\documentclass[final,12pt]{elsarticle}
\usepackage{amsmath}
\usepackage{graphicx}
\usepackage[labelformat=simple]{subcaption}  

\usepackage{amssymb}
\usepackage{amsmath}
\usepackage{graphicx}
\usepackage{subcaption}
\usepackage{booktabs}
\usepackage{xcolor}
\usepackage{ulem}
\usepackage{appendix} 
\usepackage{cleveref}
\newtheorem{proposition}{Proposition}

\journal{XXX}

\DeclareUnicodeCharacter{03D5}{$\phi$}
\DeclareUnicodeCharacter{03BE}{$\xi$}
\begin{document}

\begin{frontmatter}



\title{Localized wave solutions of three-component defocusing Kundu-Eckhaus equation with $4\times4$ matrix spectral problem}


\author[1]{Yanan Wang}
\author[2]{Min Xue\corref{cor1}}  
\ead{xuemincumtb@163.com}
\cortext[cor1]{Corresponding author}

\affiliation[1]{organization={School of Mathematical Science},
            addressline={Beihang University}, 
            city={Beijing},
            postcode={102206}, 
            country={China}}
\affiliation[2]{organization={School of Science},
            addressline={China University of Mining and Technology}, 
         city={Beijing},
          postcode={100083}, 
           country={China}}


%

\begin{abstract}
This work focuses on three-component defocusing Kundu-Eckhaus equation, which serves as a significant coupled model for describing complex wave propagation in nonlinear optical fibers. By employing binary Darboux transformation based on $4\times4$ matrix spectral problem, we derive vector dark soliton solutions, and meanwhile, the exact expressions of asymptotic dark soliton components are obtained through an asymptotic analysis method. Furthermore, breather and Y-shaped breather solutions, absent from single-component defocusing kundu-Eckhaus systems, are obtained due to the mutual coupling effects between different components. The results significantly advance our understanding nonlinear wave phenomenon induced by coupling effects and provide a theoretical reference for subsequent studies on defocusing multi-component systems.
\end{abstract}




\end{frontmatter}

\section{Introduction}
In recent years, there has been growing interest in multi-component nonlinear systems due to their crucial roles in the fields of nonlinear optical fibers, plasma astrophysics, molecular dynamics and Bose-Einstein condensates \cite{bj1,bj2,bj3,bj4,bj5}. Intrinsic coupling effects in multi-component systems give rise to richer dynamics and novel phenomena exceeding those in uncoupled systems, such as cross-phase modulation, inter-component energy transfer and bound states with relative velocities \cite{cross,vector,bound}. 

As a generalization of the nonlinear Schr\"{o}dinger (NLS) equation, it's well-known that the Kundu-Eckhaus (KE) equation \cite{kundu,Eck}
\begin{align}
\mathrm{i}q_t+q_{xx}+2\sigma|q|^2q+\delta^2|q|^4q+2\mathrm{i}\delta(|q|^2)_xq=0,
\end{align}
describes the propagation of the ultrashort optical pulses due to the non-Kerr nonlinear effect and the self-frequency shift effect. $q$ is a complex function denoting the electromagnetic wave. The parameters $\sigma$ and $\delta^2$ are the self-phase modulation coefficient and  the quintic nonlinearity coefficient, respectively. Furthermore, $\sigma=1$ corresponds to the focusing KE equation, and $\sigma=-1$ corresponds to the defocusing KE equation. Numerous methods have been employed to investigate this equation, including Darboux transformation (DT) \cite{dt1,dt2}, Hirota bilinear method \cite{hi}, tan-expansion method \cite{tan} and Riemann-Hilbert approach \cite{rh}. 

In this paper, we focus on the defocusing case of the following three-component KE (TCKE) equation,
\begin{align}\label{eq}  \mathrm{i}q_{j,t}&+q_{j,xx}+2\sigma(\sum_{k=1}^{3}|q_k|^2)q_j+\rho^2(\sum_{k=1}^{3}|q_k|^2)^2q_j+2\mathrm{i}\rho[(\sum_{k=1}^{3}|q_k|^2)q_j]_t \notag\\
&-2\mathrm{i}\rho(\sum_{k=1}^{3}q_k^*q_{k,x})q_j=0, (j=1,2,3),
\end{align}
where the asterisk denotes the complex conjugation, $\rho$ is a real parameter and $\rho^2$ is the quintic-nonlinearity coefficient. When $\sigma=-1$, we call it the defocusing TCKE equation which is regarded as an integrable extension of the classic defocusing KE equation.

The multi-component KE equations have attracted much attention. It's known that two-component KE (CKE) equation, introduced in \cite{2ke}, has been the subject of extensive research. The $3\times3$ Lax pair of the focusing CKE equation was constructed, and bright-bright soliton solutions were obtained by DT method in \cite{qi}. The dark-dark soliton solutions of the defocusing CKE equation were reached by Hirota bilinear method \cite{dk}. The first-order rogue wave, breather and interaction solutions for the focusing CKE equation were derived by generalized DT in \cite{1ro}. In \cite{2ro}, higher-order rogue wave pairs of the focusing CKE equation were discussed. \cite{3ro} presented the vector rational and semi-rational rogue wave solutions for the focusing CKE equation by binary DT method. Furthermore, for the focusing TCKE equation, the $4\times4$ Lax pair had been constructed and DT was employed to obtain first- and second-order rogue wave solutions in \cite{3ke1}. The positon solutions were obtained by degenerate DT for the focusing TCKE equation in \cite{3ke2}. The Darboux-dressing method was used to derive novel solitons, breathers and rogue waves in \cite{3ke3} for the focusing TCKE equation.

To our knowledge, several localized waves and dynamic properties of the defocusing TCKE equation have not been explored yet. Therefore, inspired by \cite{zgq0}, we utilize binary DT method to obtain vector localized wave solution and analyze the relevant dynamic behaviors. This paper is organized as follows. In Section \ref{2s}, we give the binary DT in the determinant form for the defocusing TCKE equation based on $4\times4$ Lax pair. In Section \ref{3s}, vector dark soliton solutions are exhibited and the asymptotic analysis method is used to obtain asymptotic dark soliton components. In Section \ref{4s}, breather solution and Y-shaped breather solution are derived by choosing appropriate parameters. In Section \ref{5s}, the conclusion and discussion are drawn.

\section{Binary Darboux transformation for the system \eqref{eq} with $\sigma=-1$}\label{2s}
According to the Lax pair \cite{3ke1}, we can deduce the following Lax pair of the defocusing case of system \eqref{eq}.
\begin{align}\label{lax}
    \Psi_x=U\Psi,~\Psi_t=V\Psi,
\end{align}
with $U=\mathrm{i}\lambda J+\mathrm{i}JQ+\frac{1}{2}\mathrm{i}J(\rho\sum_{k=1}^3|q_k|^2)$ and
\begin{align*}
V&=2\mathrm{i}J\lambda^2+2\mathrm{i}\lambda Q+J(Q_x-Q^2)-\mathrm{i}J(\rho v_1)^2-\mathrm{i}Q(\rho v_1)+Jv_2,
\end{align*}
where $J=\mathrm{diag}(-1,1,1,1)$, $v_1=\sum_{k=1}^3|q_k|^2$, $v_2=\frac{1}{2}\rho (\sum_{k=1}^3q_kq^*_{k,x}-q^*_kq_{k,x})$ and
\[Q=
\begin{pmatrix}
    0 & \mathbf{-q}^T  \\
    \mathbf{q^*} &  0
\end{pmatrix}, ~\mathbf{q}=(q_1,q_2,q_3)^T.\]
Here $\Psi=(\psi_1,\psi_2,\psi_3,\psi_4)^T$ is the vector eigenfunction of the spectral problem \eqref{lax} and $\lambda$ is the spectral parameter. The defocusing TCKE system can be derived from the  compatibility condition $U_t-V_x+[U,V]=0$. 

We introduce a gauge transformation to convert the spectral problem \eqref{lax} into an AKNS-type spectral problem. Let 
\begin{align*}
S=\mathrm{diag}(\mathrm{e}^{-\frac{1}{2}\rho\int v_1dx},\mathrm{e}^{\frac{1}{2}\rho\int v_1dx},\mathrm{e}^{\frac{1}{2}\rho\int v_1dx},\mathrm{e}^{\frac{1}{2}\rho\int v_1dx}),
\end{align*}
and through transformations $\Psi=S\Phi$ and $q_j=u_j\mathrm{e}^{-\mathrm{i}\rho\int w_1dx},(j=1,2,3)$, the new spectral problem can be written as follows,
\begin{align}\label{lax1}
    \Phi_x=W_1\Phi,~\Phi_t=W_2\Phi,
\end{align}
with $W_1=\mathrm{i}\lambda J+\mathrm{i}JP$ and
\begin{align*}
W_2&=2\mathrm{i}J\lambda^2+2\mathrm{i}\lambda P+J(P_x-P^2)+Jw_2-\mathrm{i}J\int w_{2,t}dx,
\end{align*}
where $J=\mathrm{diag}(-1,1,1,1)$, $w_1=\sum_{k=1}^3|u_k|^2$,  $w_2=\frac{1}{2}\rho (\sum_{k=1}^3u_ku^*_{k,x}-u^*_ku_{k,x})$ and
\[P=
\begin{pmatrix}
    0 & \mathbf{-u}^T  \\
    \mathbf{u^*} &  0
\end{pmatrix}, ~\mathbf{u}=(u_1,u_2,u_3)^T.\]
Here $\Phi=(\phi_1,\phi_2,\phi_3,\phi_4)^T$ is the vector eigenfunction of the spectral problem \eqref{lax1}. The  compatibility condition $W_{1,t}-W_{2,x}+[W_1,W_2]=0$ generate the following equation. 
\begin{align}\label{eq2}    &\mathrm{i}u_{j,t}+u_{j,xx}-2(\sum_{k=1}^{3}|u_k|^2)u_j+\mathrm{i}\rho(\sum_{k=1}^3u_ku_{k,x}^*-u^*_ku_{k,x})u_1+\rho u_1\int\sum_{k=1}^3\Big(|u_k|^2\Big)_tdx\notag\\&=0, (j=1,2,3).
\end{align}
It's known that system \eqref{eq2} and defocusing system \eqref{eq} are gauge equivalent due to the introduction of the above gauge transformation. Hence, we start from system \eqref{eq2} to derive the relevant results for system \eqref{eq} with $\sigma=-1$.

Based on numerous literatures on Binary DT \cite{bdt1,bdt2,bdt3,bdt4}, the following proposition can be established.
\begin{proposition}
    Let $\Phi_{j}=(\phi_{j1},\phi_{j2},\phi_{j3},\phi_{j4})^T,(j=1,2,\cdots,N)$ be $N$ linearly independent solutions of the spectral problem \eqref{lax1} under the spectral parameters $\lambda_j,(j=1,2,\cdots,N)$, respectively. The $N$-fold binary DT for system \eqref{eq2} is given as follows.
    \begin{align*}
        \Phi[N]&=\Phi-HW^{-1}\Omega,\\
        P[N]&=P+\mathrm{i}[J,HW_{-1}H^{\dagger}],
    \end{align*}
    with $H=(\Phi_1,\Phi_2,\cdots,\Phi_N)$ and
\[W=\begin{pmatrix}
    \Omega(\Phi_1,\Phi_1) &\Omega(\Phi_1,\Phi_2) &\cdots &\Omega(\Phi_1,\Phi_N)\\
    \Omega(\Phi_2,\Phi_1) &\Omega(\Phi_2,\Phi_2) &\cdots &\Omega(\Phi_2,\Phi_N) \\
    \vdots&\vdots&\ddots&\vdots\\
    \Omega(\Phi_N,\Phi_1) &\Omega(\Phi_N,\Phi_2) &\cdots &\Omega(\Phi_N,\Phi_N)
\end{pmatrix},~\Omega=\begin{pmatrix}
    \Omega(\Phi_1,\Phi) \\ \Omega(\Phi_2,\Phi)\\\vdots \\\Omega(\Phi_N,\Phi)
\end{pmatrix},\]
where $\dagger$ denotes the  Hermitian conjugate, $\Omega(\Phi_j,\Phi_k)=\frac{\Phi_j^{\dagger}J\Phi_k}{\mathrm{i}(\lambda_k-\lambda_j^*)}$ for $\lambda\in\mathbb{C}$ and $\Omega(\Phi_j,\Phi_j)=\lim_{\lambda_k\to \lambda_j}\frac{\Phi_j^{\dagger}J\Phi_k}{\mathrm{i}(\lambda_k-\lambda_j^*)}$ for $\lambda\in\mathbb{R}$.
\end{proposition}
Hence, $N$-order solutions for system \eqref{eq} with $\sigma=-1$ is presented as follows through the above proposition and the gauge transformation.
\begin{align}\label{qn}
    q_j[N]&=\mathrm{e}^{-\mathrm{i}\rho\int\sum_{k=1}^3|u_k|^2dx}u_j[N]\notag\\&=\mathrm{e}^{-\mathrm{i}\rho\int\sum_{k=1}^3|u_k|^2dx}\left(u_j+2\mathrm{i}\dfrac{\begin{vmatrix}
        W&H_{j+1}^{\dagger}\\
        H_1&0
    \end{vmatrix}}{\begin{vmatrix}
        W
    \end{vmatrix}}\right),j=1,2,3,
\end{align}
where $H_1=(\phi_{11},\phi_{21},\cdots,\phi_{N1})$ and $H_{j+1}=(\phi_{1,j+1},\phi_{2,j+1},\cdots,\phi_{N,j+1})$.

In addition, when the spectral parameter $\lambda\in\mathbb{R}$, we need to consider the limit form of binary DT. For instance, the following one-fold binary DT is employed.  
\begin{align}\label{one}
    \Phi[1]&=\lim_{\nu\to\lambda_1}\left(I-\frac{(\nu-\lambda_1)\Phi_1\Phi_1^{\dagger}J}{(\lambda-\lambda_1)\Phi_1^{\dagger}J\Xi(\nu)}\right)\Phi,\notag\\    P[1]&=\lim_{\nu\to\lambda_1}\left(P+\left[J,\frac{(\nu-\lambda_1)\Phi_1\Phi_1^{\dagger}}{\Phi_1^{\dagger}J\Xi(\nu)}\right]\right),
\end{align}
where 
\begin{align*}
    \Xi(\nu)=\Phi_1(\nu)+\frac{\beta(\nu-\lambda_1)}{B}\widetilde{\Phi_1}(\lambda_1),~\left(\beta\neq\lim_{\nu\to\lambda_1}\frac{\Phi_1^{\dagger}J\Phi_1(\nu)}{-\nu+\lambda_1}\right),
\end{align*}
and $\widetilde{\Phi_1}$ is another spectial solution of the spectral problem \eqref{lax1} under $\lambda=\lambda_1$ which meets $\Phi_1^{\dagger}J\widetilde{\Phi_1}\equiv B=\mathrm{const}\neq0$, but $\Phi_1^{\dagger}J\Phi_1=0$. 

\section{Dark solitons and asymptotic analysis}\label{3s}
In this section, we start from the plane wave background to construct the $N$-dark soliton solutions. Initially, the plane wave seed solutions are taken as $u_j=c_j\mathrm{e}^{\mathrm{i}\alpha_j}=c_j\mathrm{e}^{\mathrm{i}(a_jx+b_jt)},j=1,2,3$ with $b_j=2\rho(\sum_{k=1}^3a_kc_k^2-a_j^2-2\sum_{k=1}^3c_k^2),j=1,2,3$, where $a_j,c_j$ are the real parameters. The dark soliton solution appears when $a_j(j=1,2,3)$ are mutually distinct. Hence, we assume $a_1\neq a_2\neq a_3$ in this section.

\begin{figure}[ht!]
    \centering
   \begin{subfigure}{0.3\textwidth}
        \centering
        \includegraphics[width=\textwidth]{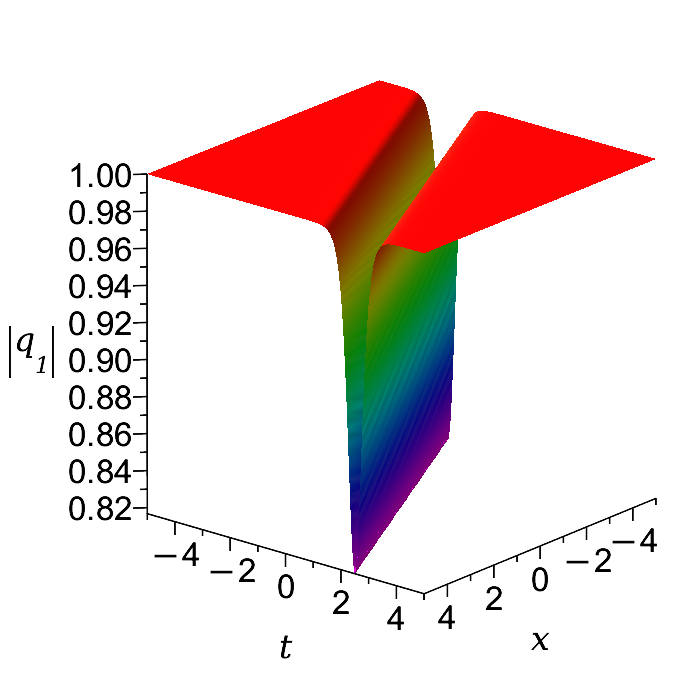}
        \caption{}
        \label{fg11}
    \end{subfigure}
    \begin{subfigure}{0.3\textwidth}
        \centering
        \includegraphics[width=\textwidth]{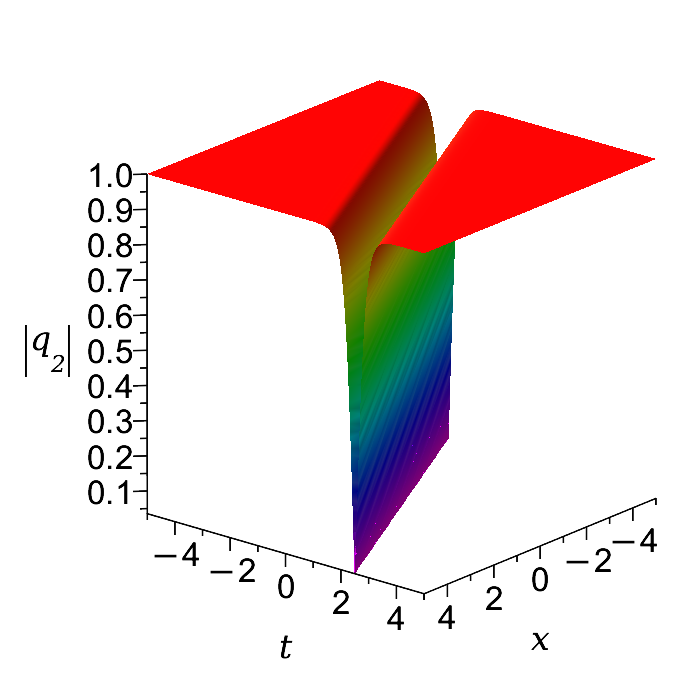}
        \caption{}
        \label{fg12}
    \end{subfigure}
    \begin{subfigure}{0.3\textwidth}
        \centering
        \includegraphics[width=\textwidth]{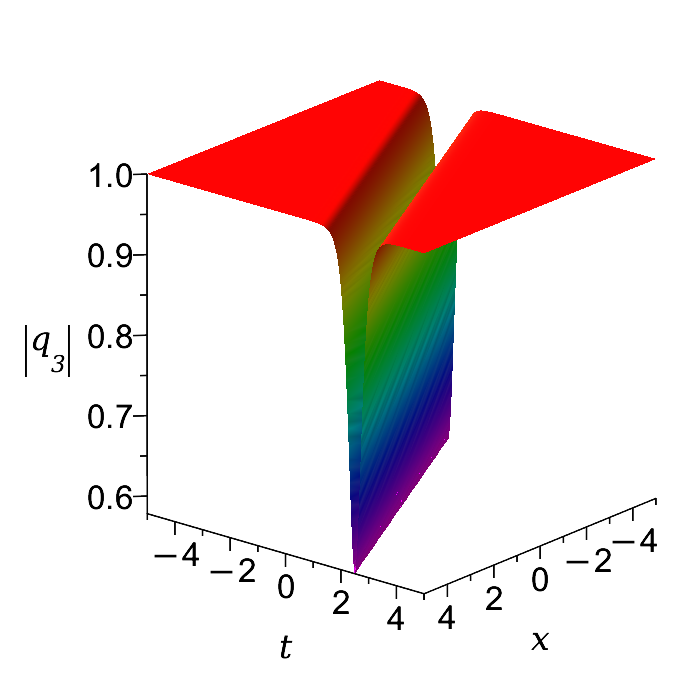}
        \caption{}
        \label{fg13}
    \end{subfigure}
     \caption{The single dark soliton solution of the system \eqref{eq} when $\sigma=-1$ with $a_1=-1,a_2=1,a_3=2,\rho=1,c_1=c_2=c_3=1,\gamma=0,\lambda_1=-\frac{1}{2}$.}
    \label{fig1}
\end{figure}

\begin{figure}[ht!]
    \centering
   \begin{subfigure}{0.3\textwidth}
        \centering
        \includegraphics[width=\textwidth]{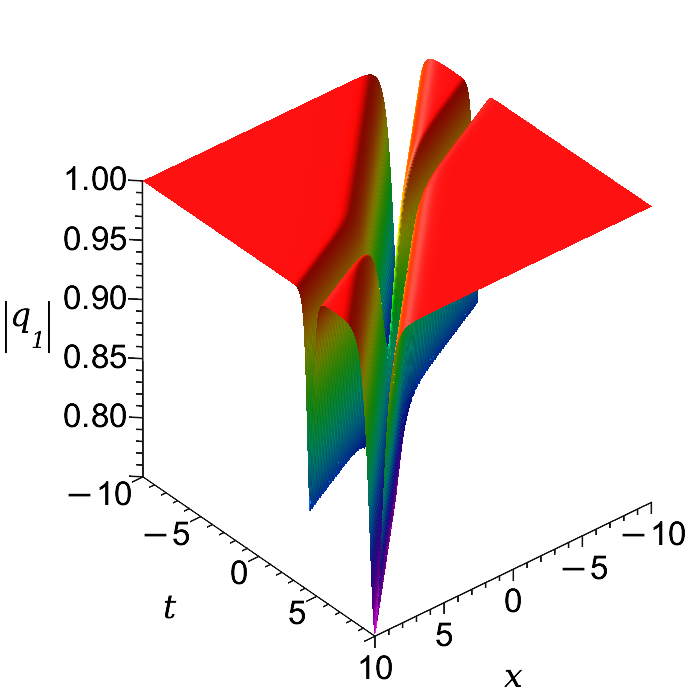}
        \caption{}
        \label{fg21}
    \end{subfigure}
    \begin{subfigure}{0.3\textwidth}
        \centering
        \includegraphics[width=\textwidth]{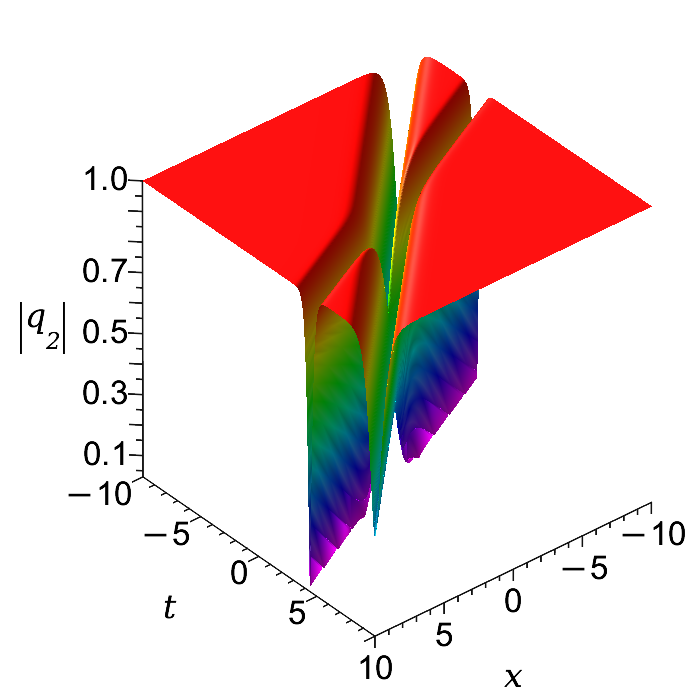}
        \caption{}
        \label{fg22}
    \end{subfigure}
    \begin{subfigure}{0.3\textwidth}
        \centering
        \includegraphics[width=\textwidth]{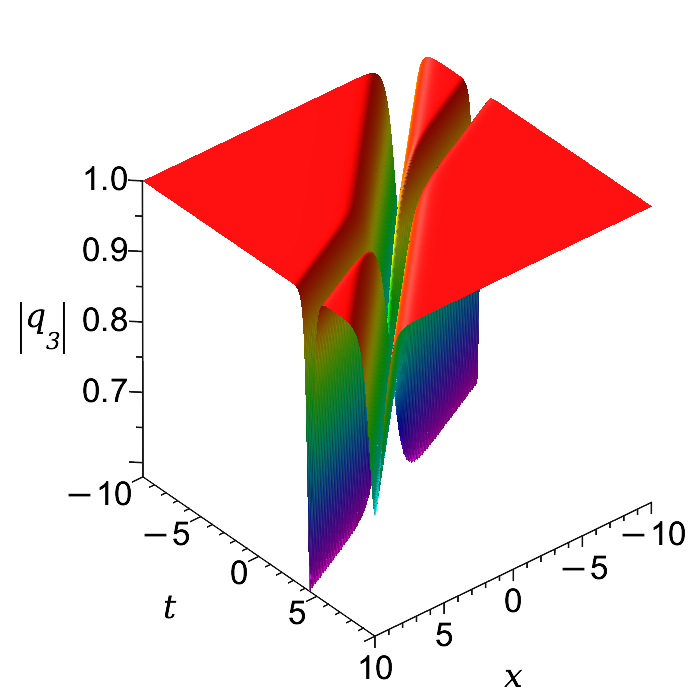}
        \caption{}
        \label{fg23}
    \end{subfigure}
     \caption{The second-order dark soliton solution of the system \eqref{eq} when $\sigma=-1$ with $a_1=-1,a_2=1,a_3=2,\rho=1,c_1=c_2=c_3=1,\gamma=0,\lambda_1=-\frac{1}{2},\lambda_2=-\frac{1}{8}$.}
    \label{fig2}
\end{figure}

Under this seed solution, the general solution for the spectral problem \eqref{lax1} is obtained with $\lambda=\lambda_k$. 
\begin{align}
     \Phi_k=\left(\mathrm{e}^{\mathrm{i}\theta_k},\frac{c_1\mathrm{e}^{\mathrm{i}(\theta_k-\alpha_1)}}{\mu_k-a_1-\lambda_k},\frac{c_2\mathrm{e}^{\mathrm{i}(\theta_k-\alpha_2)}}{\mu_k-a_2-\lambda_k},\frac{c_3\mathrm{e}^{\mathrm{i}(\theta_k-\alpha_3)}}{\mu_k-a_3-\lambda_k}\right)^T,
\end{align}
with $\theta_k=\mu_kx+(-\mu_k^2+2\lambda_k\mu_k+\rho\sum_{n=1}^3a_nc_n^2-2\sum_{n=1}^3c_n^2+\lambda_k^2)t$, where $\mu_k$ is the root of the following characteristic equation,
\begin{align}\label{ce}
    \lambda_k+\mu_k+\sum_{n=1}^3\frac{c_k^2}{\mu_k-a_n-\lambda_k}=0,k=1,2,\cdots,N.
\end{align}
When $\lambda\in\mathbb{R}$, we assume $\mu_j=\mathrm{Re}(\mu_{j})+\mathrm{i}\mathrm{Im}(\mu_{j})$  and $\beta=\frac{2\mathrm{e}^{2\gamma\mathrm{Im}(\mu_{j})}}{\mu_j-\mu_j^*}(\gamma\in\mathbb{R})$. Then by a direct symbol computation, we can derive 
\begin{align*}
\Omega(\Phi_j,\Phi_k)&=\frac{2\mathrm{ie}^{\mathrm{i}(\theta_k+\theta_j^*)}}{\lambda_j-\lambda_k+\mu_k-\mu_j},(j\neq k), \\
    \Omega(\Phi_j,\Phi_j)&=\lim_{\lambda_k\to\lambda_j}\Omega(\Phi_j,\Phi_k)=\frac{2\mathrm{i}[\mathrm{e}^{\mathrm{i}(\theta_j+\theta_j^*)}+\mathrm{e}^{2\gamma\mathrm{Im}{(\mu_{1})}}]}{\mu_j^*-\mu_j}.
\end{align*}
From \eqref{one}, the explict one-fold binary DT matrix is given,
\begin{align*}
 T=I+\frac{(\mu_1^*-\mu_1)\Phi_{1}\Phi_1^{\dagger}J}{2(\lambda-\lambda_1)(\mathrm{e}^{\mathrm{i}(\theta_j+\theta_j^*)}+\mathrm{e}^{2\gamma\mathrm{Im}{(\mu_{1})}})}.
\end{align*}
Therefore, the vector single dark soliton solutions for the system \eqref{eq2} are exhibited as
\begin{align}\label{exact}
    u_j[1]=c_j\mathrm{e}^{\mathrm{i}(a_jx+b_jt)}\{1-R_j-R_j\tanh{[\mathrm{Im}(\mu_{1})(2\mathrm{Re}(\mu_{1})t-2\lambda_1t-x})]\},
\end{align}
where $R_j=\frac{\mathrm{i}\mathrm{Im}(\mu_{1})}{\mathrm{i}\mathrm{Im}(\mu_{1})+a_j-\mathrm{Re}(\mu_{1})+\lambda_1},j=1,2,3$, and then the vector single dark soliton solutions for the system \eqref{eq} with $\sigma=-1$ is $q_j=u_j\mathrm{e}^{-\mathrm{i}\rho\int w_1dx},j=1,2,3$ with the velocity $\frac{\mathrm{Im}(\mu_{1})\mathrm{Re}(\mu_{1})-\mathrm{Im}(\mu_{1})\lambda_1}{\mathrm{Im}(\mu_{1})}$. 
The vector single dark soliton solution of the system \eqref{eq} with $\sigma=-1$ is shown in Fig.\ref{fig1} by choosing appropriate parameters.
According to \eqref{qn}, when $N=2$, the second-order vector dark soliton solution is constructed as presented in Fig.\ref{fig2}. 

Subsequently, the asymptotic analysis method described in \cite{dt2} is employed to derive the asymptotic dark solitons and dynamical properties for the second-order vector dark soliton solution. 

For the above second-order solution, we set $\widetilde{\theta_j}=\mathrm{Im}(\mu_{j})[2\mathrm{Re}(\mu_{j})t-2\lambda_jt-x],j=1,2$.
Because of $\lambda_1<\lambda_2$, when $t\to\pm\infty$, along the characteristic line $x-2\mathrm{Re}(\mu_{1})t+2\lambda_1t=\mathrm{const}$, we have $\widetilde{\theta_2}\to\pm\infty$. Hence, we perform the following elementary transformation of the determinant under $N=2$.
\begin{align}\label{D1}
    \begin{vmatrix}
        W&H_{j+1}^{\dagger}\\
        H_1&0
    \end{vmatrix}=&\begin{vmatrix}
        \Omega(\Phi_1,\Phi_1)&\Omega(\Phi_1,\Phi_2)&\phi_{1,j+1}^*\\
        \Omega(\Phi_2,\Phi_1)&\Omega(\Phi_2,\Phi_2)&\phi_{2,j+1}^*\\
        \phi_{11}&\phi_{21}&0
    \end{vmatrix},j=1,2,3,\\
  \to&
  \left\{ 
    \begin{aligned}
        &\begin{vmatrix}
        \Omega(\Phi_1,\Phi_1)&\Delta_{12}\mathrm{e}^{\widetilde{\theta_1}+\mathrm{i}r_1}&\phi_{1,j+1}^*\\
        \Delta_{21}\mathrm{e}^{\widetilde{\theta_1}+\mathrm{i}r_2}&\Delta_{22}&(\phi_{2,j+1}\mathrm{e}^{-\widetilde{\theta_2}})^*\\
        \phi_{11}&\phi_{21}\mathrm{e}^{-\widetilde{\theta_2}}&0
    \end{vmatrix}  \quad t\to+\infty, \\
        &\begin{vmatrix}
        \Omega(\Phi_1,\Phi_1)&0&\phi_{1,j+1}^*\\
        0&\Delta_{22}&0\\
        \phi_{11}&0&0
    \end{vmatrix} \quad t\to-\infty,
    \end{aligned}
    \right.
\end{align}
where $r_1,r_2$ are both functions on $x$ and $t$. When $t\to\pm\infty$, along the characteristic line $x-2\mathrm{Re}(\mu_{2})t+2\lambda_2t=\mathrm{const}$, we have $\widetilde{\theta_1}\to\pm\infty$. Hence, we obtain the following results.
\begin{align}\label{D2}
    \begin{vmatrix}
        W&H_{j+1}^{\dagger}\\
        H_1&0
    \end{vmatrix}=&\begin{vmatrix}
        \Omega(\Phi_1,\Phi_1)&\Omega(\Phi_1,\Phi_2)&\phi_{1,j+1}^*\\
        \Omega(\Phi_2,\Phi_1)&\Omega(\Phi_2,\Phi_2)&\phi_{2,j+1}^*\\
        \phi_{11}&\phi_{21}&0
    \end{vmatrix},j=1,2,3,\\
  \to&
  \left\{ 
    \begin{aligned}
        &\begin{vmatrix}
        \Delta_{11}&\Delta_{12}\mathrm{e}^{\widetilde{\theta_2}+\mathrm{i}r_1}&(\phi_{1,j+1}\mathrm{e}^{-\widetilde{\theta_1}})^*\\
        \Delta_{21}\mathrm{e}^{\widetilde{\theta_2}+\mathrm{i}r_2}& \Omega(\Phi_2,\Phi_2)&\phi_{2,j+1}^*\\
        \phi_{11}\mathrm{e}^{-\widetilde{\theta_1}}&\phi_{21}&0
    \end{vmatrix}  \quad t\to+\infty, \\
        &\begin{vmatrix}
       \Delta_{11}&0&0\\
        0&\Omega(\Phi_2,\Phi_2)&\phi_{2,j+1}^*\\
        0&\phi_{21}&0
    \end{vmatrix} \quad t\to-\infty.
    \end{aligned}
    \right.
\end{align}

In summary, we substitute \eqref{D1} and \eqref{D2} into \eqref{qn} so that asymptotic vector dark soliton components for the second-order vector dark soliton are derived. Namely, $|[q_j]^{\pm}|^2=|[u_j]^{\pm}|^2,j=1,2,3$ and when $t\to\pm\infty$,
\begin{align*}
  [u_j]^-_\mathrm{I} &\to \mathrm{e}^{\mathrm{i}(a_jx+b_jt)}\left\{A_{j1}-A_{j2}\tanh\left[\sqrt{2}(x-2t)\right]\right\}, \\
    [u_j]^+_\mathrm{I}&\to \mathrm{e}^{\mathrm{i}(a_jx+b_jt)}\left\{B_{j1}-B_{j2}\tanh\left[\sqrt{2}(x-2t)+\frac{1}{2}\ln\left(\frac{2\sqrt{14}-7}{2\sqrt{14}+7}\right)\right]\right\},\\ 
    [u_j]^-_\mathrm{II}&\to \mathrm{e}^{\mathrm{i}(a_jx+b_jt)}\left\{C_{j1}-C_{j2}\tanh\left[\frac{\sqrt{7}}{2}(x-t)+\frac{1}{2}\ln\left(\frac{2\sqrt{14}-7}{2\sqrt{14}+7}\right)\right]\right\},\\
    [u_j]^+_\mathrm{II} &\to \mathrm{e}^{\mathrm{i}(a_jx+b_jt)}\left\{D_{j1}-D_{j2}\tanh\left[\frac{\sqrt{7}}{2}(x-t)\right]\right\},
\end{align*}
where $A_{jk},B_{jk},C_{jk},D_{jk}(j=1,2,3,k=1,2)$ are displayed in the Appendix.

As observed in Fig.\ref{fg3}, the first dark soliton component $q_1$ contains two asymptotic dark soliton components $[q_1]_{\mathrm{I}},[q_1]_{\mathrm{II}}$ which consist of four expressions $[q_1]_{\mathrm{I}}^{\pm}, [q_1]_{\mathrm{II}}^{\pm}$. We also find the phases of the two solitons change during the collision process so that they maintain an elastic interaction. In addition, it is necessary to verify the validity of the asymptotic solitons. In Fig.\ref{fg1} and Fig.\ref{fg2}, the four asymptotic dark soliton components (blue and red dotted line) of $q_1$ match the exact solution (green line) perfectly in the far-field region. The cases of $q_2$ and $q_3$ are similar to that of $q_1$.
\begin{figure}[ht!]
    \centering
   \begin{subfigure}{0.31\textwidth}
        \centering
        \includegraphics[width=\textwidth]{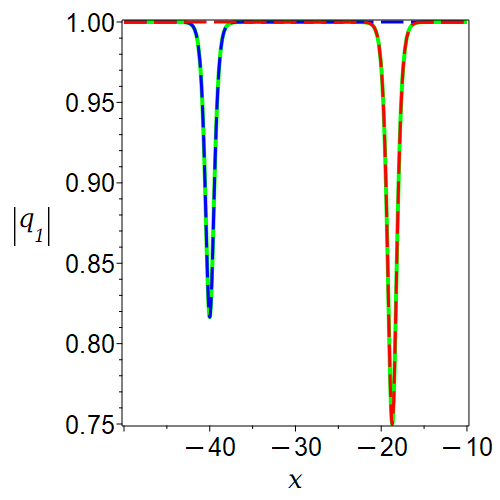}
        \caption{$t=-20$}
        \label{fg1}
    \end{subfigure}
    \hspace{0.2cm}
    \begin{subfigure}{0.31\textwidth}
        \centering
        \includegraphics[width=\textwidth]{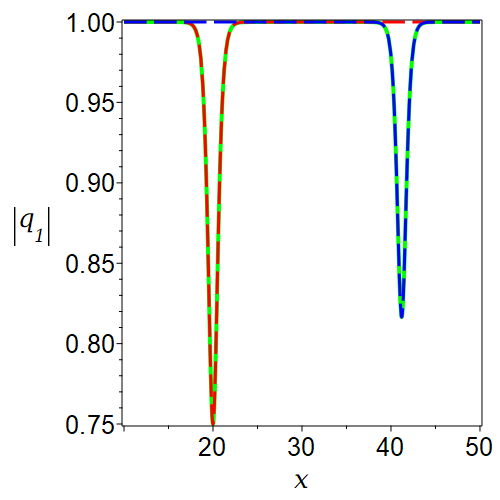}
        \caption{$t=20$}
        \label{fg2}
    \end{subfigure}
        \hspace{0.2cm}
    \begin{subfigure}{0.3\textwidth}
        \centering
        \includegraphics[width=\textwidth]{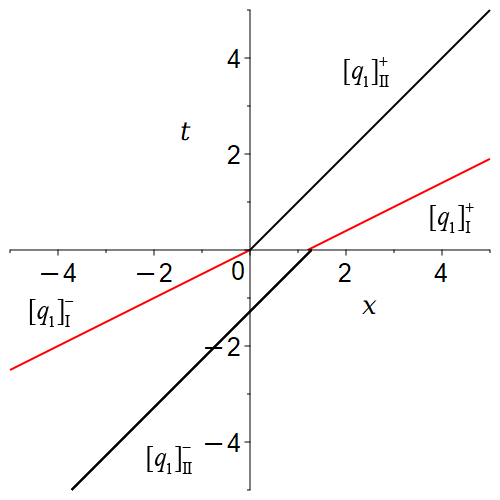}
        \caption{}
        \label{fg3}
    \end{subfigure}
     \caption{(a)(b) The comparison of the asymptotic solitons and the exact solution for the dark soliton component $q_1$. Blue: asymptotic soliton $[q_1]_{\mathrm{I}}$; Red: asymptotic soliton $[q_1]_{\mathrm{II}}$; Green: the exact solution \eqref{exact}; (c) The wave crest line graph of the asymptotic solitons for the dark soliton component $q_1$.}
    \label{fig}
\end{figure}

\section{Breather solutions}\label{4s}
In this section, we further explore the breather solutions for the system \eqref{eq} with $\sigma=-1$ when $\lambda\in\mathbb{C}$. This feature is absent from the defocusing single-component KE equation.

Firstly, we start with $a_1=a_2=a_3$ so that the characteristic equation \eqref{ce} has three distinct complex roots, corresponding to $\mu_1=\frac{a_1}{2}+\frac{\chi_1}{2}$, $\mu_2=\frac{a_1}{2}-\frac{\chi_1}{2}$ and $\mu_3=a_1+\lambda_1$ where $\chi_1=\sqrt{a_1^2+4\lambda_1a_1-4\sum_{k=1}^3c_k^2+4\lambda_1^2}$. In other words, the matrix $-\mathrm{i}W_1$ can be diagonalized by $ P_1^{-1}(-\mathrm{i}W_1)P_1=\Lambda$, where
\[P_1=\begin{pmatrix}
    -\frac{\sum_{k=1}^3c_k^2}{c_3(\mu_1+\lambda_1)} &  -\frac{\sum_{k=1}^3c_k^2}{c_3(\mu_2+\lambda_1)}&0&0\\
    \frac{c_1}{c_3}&\frac{c_1}{c_3}&-\frac{c_3}{c_1}&-\frac{c_2}{c_1}  \\
    \frac{c_2}{c_3}&\frac{c_2}{c_3}&0&1\\
    1&1&1&0
\end{pmatrix},\Lambda=\begin{pmatrix}
    \mu_1&0&0&0\\0&\mu_2&0&0\\0&0&a_1+\lambda_1&0\\0&0&0&a_1+\lambda_1
\end{pmatrix}.\]
Through a direct calculation, it's known that the general solution for the spectral problem \eqref{lax1} can be generated by $\Phi_1=G^{-1}P_1Nl$, where $G=\mathrm{diag}(1,\mathrm{e}^{\hat{\alpha}_1},\mathrm{e}^{\hat{\alpha}_2},\mathrm{e}^{\hat{\alpha}_3})$, $l=(l_1,l_2,l_3,l_4)^T$ and $N=\mathrm{diag}(\mathrm{e}^{\zeta(\mu_1)},\mathrm{e}^{\zeta(\mu_2)},\mathrm{e}^{\zeta(a_1+\lambda_1)},\mathrm{e}^{\zeta(a_1+\lambda_1)})$ where $\hat{\alpha}_j=\mathrm{i}(a_1x+b_jt),j=1,2,3$ and $\zeta(\mu)=\mathrm{i}[\mu x+(-\mu^2+2\lambda_1\mu+\rho\sum_{k=1}^3a_1c_k^2-2\sum_{k=1}^3c_k^2+\lambda_1^2)t]$. Hence, we have
\begin{align*}
   \Omega(\Phi_1,\Phi_1)=\frac{\Phi_1^{\dagger}J\Phi_1}{\mathrm{i}(\lambda_1-\lambda_1^*)}=\frac{(G^{-1}P_1Nl)^{\dagger}JG^{-1}P_1Nl}{\mathrm{i}(\lambda_1-\lambda_1^*)}=\frac{(Nl)^{\dagger}(P_1^{\dagger}JP_1)Nl}{\mathrm{i}(\lambda_1-\lambda_1^*)}.
\end{align*}
To simplify the calculation and make $\Omega(\Phi_1,\Phi_1)\neq0$ hold identically, we transform it into a standard quadratic form based on the theory of matrix congruence. According to the characteristic equation \eqref{ce} with $a_1=a_2=a_3$, we can obtain
\[
    P_1^{\dagger}JP_1=\begin{pmatrix}
        \frac{\sum_{k=1}^3c_k^2}{c_3^2}\frac{2(\lambda-\lambda^*)}{\lambda-\lambda^*+\mu_1-\mu_1^*} &\frac{\sum_{k=1}^3c_k^2}{c_3^2}\frac{2(\lambda-\lambda^*)}{\lambda-\lambda^*+\mu_2-\mu_1^*} &0&0\\
        \frac{\sum_{k=1}^3c_k^2}{c_3^2}\frac{2(\lambda-\lambda^*)}{\lambda-\lambda^*+\mu_1-\mu_2^*} &\frac{\sum_{k=1}^3c_k^2}{c_3^2}\frac{2(\lambda-\lambda^*)}{\lambda-\lambda^*+\mu_2-\mu_2^*} &0&0\\
        0&0&1+\frac{c_3^2}{c_1^2}&\frac{c_2c_3}{c_1^2}\\0&0&\frac{c_2c_3}{c_1^2}&1+\frac{c_2^2}{c_1^2}
    \end{pmatrix}.
\]
According to the lemma in \cite{zgq}, by considering the leading principal minor of $P_1^{\dagger}JP_1$, we can derive 
\begin{align*}
    P_1^{\dagger}JP_1\simeq\mathrm{diag}(\frac{\mathrm{Im}(\lambda_1)}{\mathrm{Im}(\lambda_1)+\mathrm{Im}(\mu_1)},\frac{\mathrm{Im}(\lambda_1)}{\mathrm{Im}(\lambda_1)+\mathrm{Im}(\mu_2)},1,1).
\end{align*}
We assume $\mathrm{Im}(\lambda_1)>0$, $\mathrm{Im}(\lambda_1)+\mathrm{Im}(\mu_1)>0$ and $\mathrm{Im}(\lambda_1)+\mathrm{Im}(\mu_2)<0$, then only if $l_2=0,l_j\neq0,j=1,3,4$, the condition $\Omega(\Phi_1,\Phi_1)\neq0$ holds  identically. 

\begin{figure}[ht!]
    \centering
   \begin{subfigure}{0.3\textwidth}
        \centering
        \includegraphics[width=\textwidth]{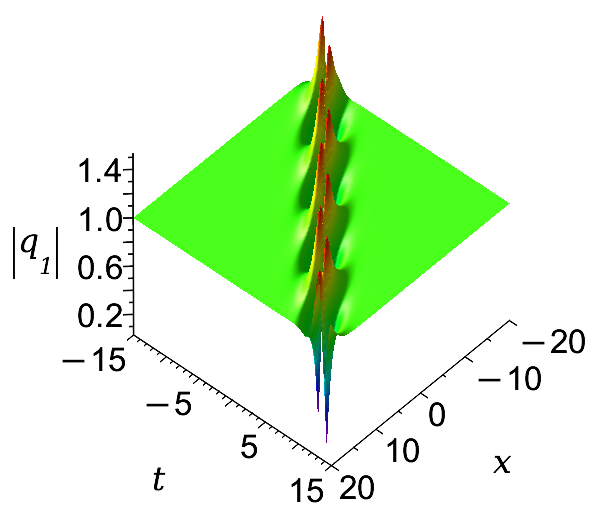}
        \caption{}
        \label{fg31}
    \end{subfigure}
    \begin{subfigure}{0.3\textwidth}
        \centering
        \includegraphics[width=\textwidth]{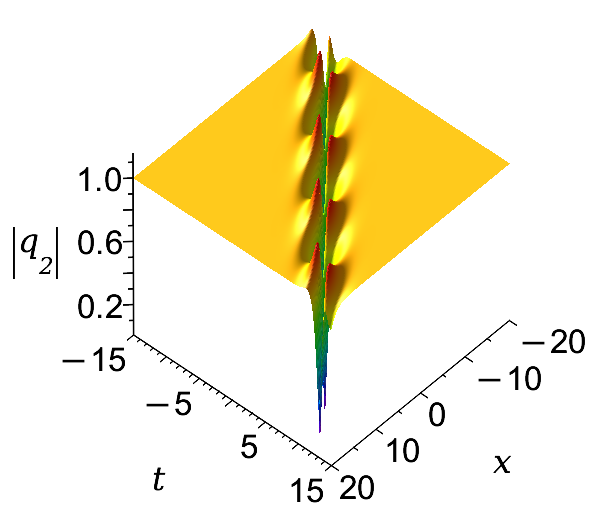}
        \caption{}
        \label{fg32}
    \end{subfigure}
    \begin{subfigure}{0.3\textwidth}
        \centering
        \includegraphics[width=\textwidth]{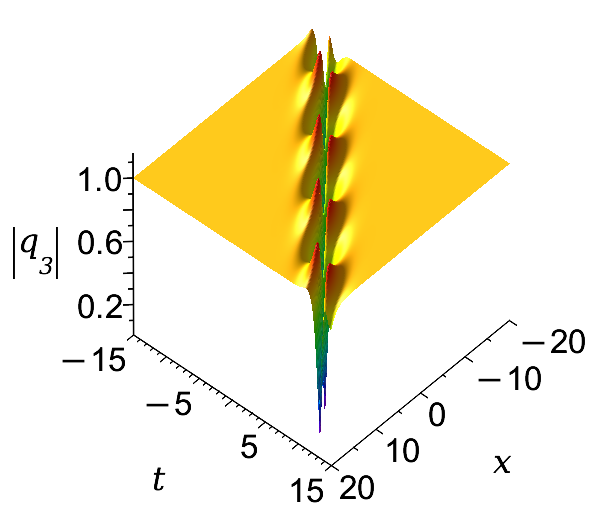}
        \caption{}
        \label{fg33}
    \end{subfigure}
     \caption{The breather solution for the system \eqref{eq} when $\sigma=-1$ with $a_1=a_2=a_3=1,\rho=1,c_1=c_2=c_3=1,\lambda_1=\mathrm{i}$.}
    \label{fig3}
\end{figure}

Therefore, based on \eqref{qn}, the first-order breather solutions for the system \eqref{eq2} and \eqref{eq} with $\sigma=-1$ are  
\begin{align*}
    u_j[1]&=c_j\mathrm{e}^{\mathrm{i}(a_1x+b_jt)}\left\{1+E_j\left[1+
    \tanh\left(\frac{K_1}{2}+\frac{1}{2}\ln K_2\right)+F_j\mathrm{e}^{\mathrm{i}K_3}\mathrm{sech}\left(\frac{K_1}{2}+\frac{1}{2}\ln K_2\right)\right]\right\},\\
    q_j[1]&=u_j[1]\mathrm{e}^{-\mathrm{i}\rho\int\sum_{k=1}^3|u_j[1]|^2dx},j=1,2,3,
\end{align*}
where 
\begin{align*}
    E_1&=E_2=E_3=\frac{\mathrm{Im}(\lambda_1)+\mathrm{Im}(\mu_1)}{K_4},F_1=-\frac{[\mathrm{Im}(\lambda_1)+\mathrm{Im}(\mu_1)]c_3(c_2l_4+c_3l_3)}{c_1^2K_4},\\F_2&=\frac{[\mathrm{Im}(\lambda_1)+\mathrm{Im}(\mu_1)]c_3}{c_2K_4},F_3=\frac{[\mathrm{Im}(\lambda_1)+\mathrm{Im}(\mu_1)]l_3}{K_4},\\
    K_4&=\mathrm{i}\mathrm{Re}(\lambda_1)+\mathrm{i}\mathrm{Re}(\mu_1)-\mathrm{Im}(\lambda_1)-\mathrm{Im}(\mu_1),\\
    K_1&=[(2a_1+8\mathrm{Re}(\lambda_1)-4\mathrm{Re}(\mu_1))\mathrm{Im}(\lambda_1)+2(a_1-2\mathrm{Re}(\lambda_1))\mathrm{Im}(\mu_1)]t\\&\quad+2(\mathrm{Im}(\lambda_1)-\mathrm{Im}(\mu_1))x,\\
    K_2&=\frac{2\mathrm{Im}(\lambda_1)c_1^2(c_1^2+c_2^2+c_3^2)}{(c_3^2l_3^2+2c_2c_3l_3l_4+c_2^2l_4^2+c_1^2l_3^2+c_1^2l_4^2)c_3^2(\mathrm{Im}(\lambda_1)+\mathrm{Im}(\mu_1))},\\
    K_3&=[a_1^2-(\mathrm{Re}(\lambda_1)+\mathrm{Re}(\mu_1))a_1-2\mathrm{Im}(\lambda_1)\mathrm{Im}(\mu_1)+\sum_{k=1}^3c_k^2-2\mathrm{Re}(\lambda_1)^2\\&\quad+2\mathrm{Re}(\lambda_1)\mathrm{Re}(\mu_1)+2\mathrm{Im}(\lambda_1)^2]t-(a_1+\mathrm{Re}(\lambda_1)-\mathrm{Re}(\mu_1))x.
\end{align*}
When we choose the appropriate parameters, the breather is shown in Fig.\ref{fig3}.

Subsequently, we consider the breather solution when $a_1\neq a_2\neq a_3$ and $\lambda\in\mathbb{C}$. The characteristic equation \eqref{ce} has four distinct complex roots, which we denote as $\hat{\mu}_1$, $\hat{\mu}_2$, $\hat{\mu}_3$ and $\hat{\mu}_4$. By a similar symbol computation, we obtain 
\[\hat{P}_1^{\dagger}J\hat{P}_1=-\begin{pmatrix}
    \frac{2(\lambda_1-\lambda_1^*)}{\lambda_1-\lambda_1^*+\hat{\mu}_1^*-\hat{\mu}_1}&  \frac{2(\lambda_1-\lambda_1^*)}{\lambda_1-\lambda_1^*+\hat{\mu}_1^*-\hat{\mu}_2}&\frac{2(\lambda_1-\lambda_1^*)}{\lambda_1-\lambda_1^*+\hat{\mu}_1^*-\hat{\mu}_3}&\frac{2(\lambda_1-\lambda_1^*)}{\lambda_1-\lambda_1^*+\hat{\mu}_1^*-\hat{\mu}_4}\\
   \frac{2(\lambda_1-\lambda_1^*)}{\lambda_1-\lambda_1^*+\hat{\mu}_2^*-\hat{\mu}_1}&\frac{2(\lambda_1-\lambda_1^*)}{\lambda_1-\lambda_1^*+\hat{\mu}_2^*-\hat{\mu}_2}&\frac{2(\lambda_1-\lambda_1^*)}{\lambda_1-\lambda_1^*+\hat{\mu}_2^*-\hat{\mu}_3} &\frac{2(\lambda_1-\lambda_1^*)}{\lambda_1-\lambda_1^*+\hat{\mu}_2^*-\hat{\mu}_4}\\
    \frac{2(\lambda_1-\lambda_1^*)}{\lambda_1-\lambda_1^*+\hat{\mu}_3^*-\hat{\mu}_1}&\frac{2(\lambda_1-\lambda_1^*)}{\lambda_1-\lambda_1^*+\hat{\mu}_3^*-\hat{\mu}_2}&\frac{2(\lambda_1-\lambda_1^*)}{\lambda_1-\lambda_1^*+\hat{\mu}_3^*-\hat{\mu}_3}&\frac{2(\lambda_1-\lambda_1^*)}{\lambda_1-\lambda_1^*+\hat{\mu}_3^*-\hat{\mu}_4}\\
    \frac{2(\lambda_1-\lambda_1^*)}{\lambda_1-\lambda_1^*+\hat{\mu}_4^*-\hat{\mu}_1}&\frac{2(\lambda_1-\lambda_1^*)}{\lambda_1-\lambda_1^*+\hat{\mu}_4^*-\hat{\mu}_2}&\frac{2(\lambda_1-\lambda_1^*)}{\lambda_1-\lambda_1^*+\hat{\mu}_4^*-\hat{\mu}_3}&\frac{2(\lambda_1-\lambda_1^*)}{\lambda_1-\lambda_1^*+\hat{\mu}_4^*-\hat{\mu}_4}
\end{pmatrix},\]
and $\hat{P}_1^{\dagger}J\hat{P}_1\simeq\mathrm{diag}(\frac{2\mathrm{Im}(\lambda_1)}{\mathrm{Im}(\lambda_1)-\mathrm{Im}(\hat{\mu}_1)},\frac{2\mathrm{Im}(\lambda_1)}{\mathrm{Im}(\lambda_1)-\mathrm{Im}(\hat{\mu}_2)},\frac{2\mathrm{Im}(\lambda_1)}{\mathrm{Im}(\lambda_1)-\mathrm{Im}(\hat{\mu}_3)},\frac{2\mathrm{Im}(\lambda_1)}{\mathrm{Im}(\lambda_1)-\mathrm{Im}(\hat{\mu}_4)})$.

\begin{figure}[ht!]
    \centering
   \begin{subfigure}{0.32\textwidth}
        \centering
        \includegraphics[width=\textwidth]{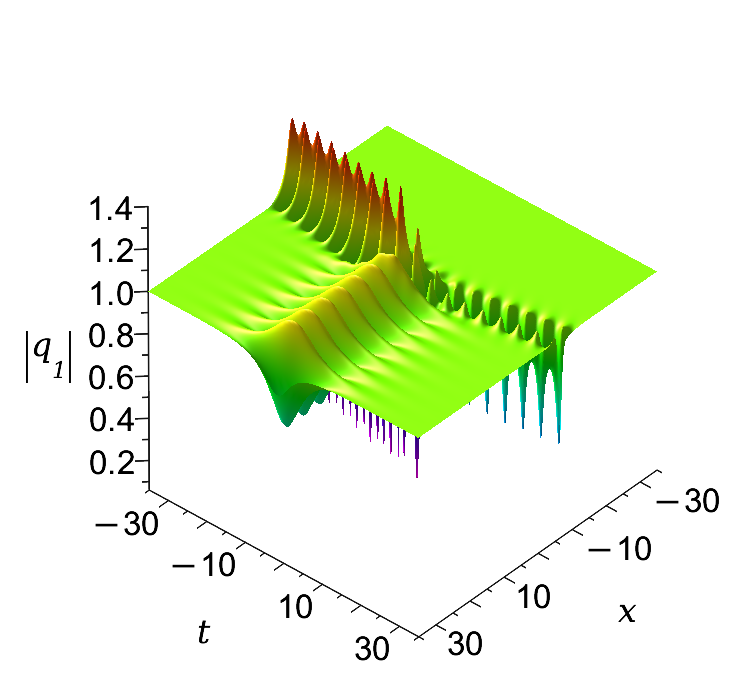}
        \caption{}
        \label{fg41}
    \end{subfigure}
    \begin{subfigure}{0.32\textwidth}
        \centering
        \includegraphics[width=\textwidth]{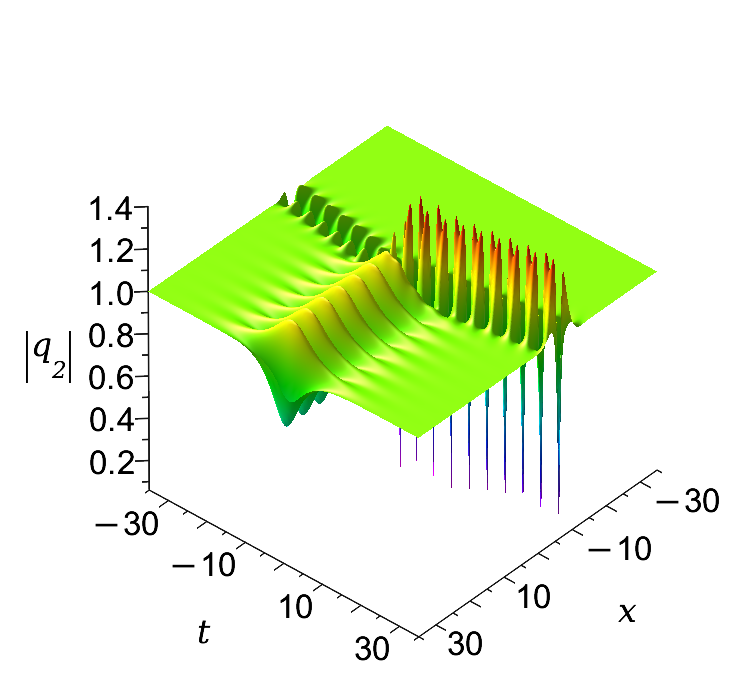}
        \caption{}
        \label{fg42}
    \end{subfigure}
    \begin{subfigure}{0.32\textwidth}
        \centering
        \includegraphics[width=\textwidth]{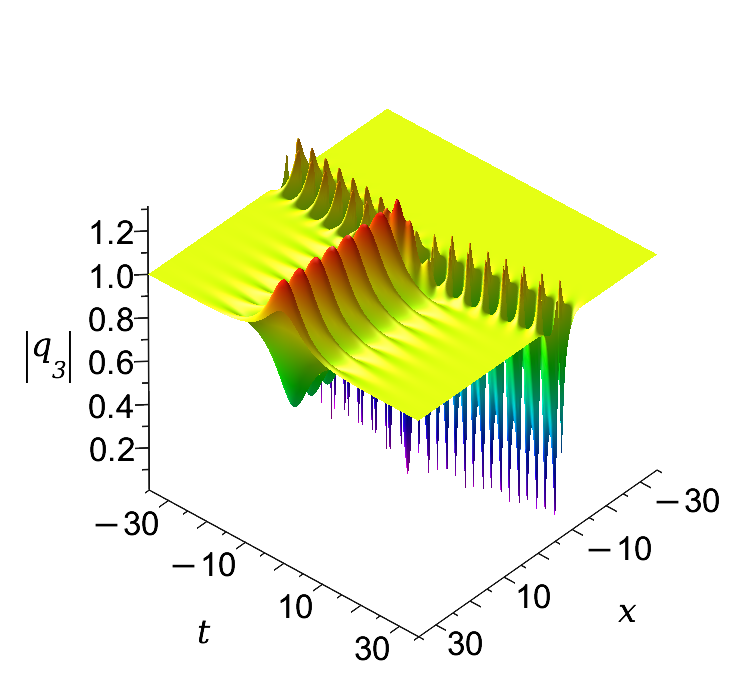}
        \caption{}
        \label{fg43}
    \end{subfigure}
     \caption{The Y-shaped breather solution for the system \eqref{eq} when $\sigma=-1$ with $a_1=-1,a_2=1,a_3=0,\rho=1,c_1=c_2=c_3=1,\lambda_1=\frac{\mathrm{i}}{2}$.}
    \label{fig4}
\end{figure}

Hence, we assume $\mathrm{Im}(\lambda_1)>0$, $\mathrm{Im}(\lambda_1)-\mathrm{Im}(\hat{\mu}_j)>0,j=1,2,4$ and $\mathrm{Im}(\lambda_1)-\mathrm{Im}(\hat{\mu}_3)<0$, then only if $l_3=0,l_j\neq0(j=1,2,4)$, the condition $\Omega(\Phi_1,\Phi_1)\neq0$ holds  identically. Finally, based on \eqref{qn}, we can obtain the first-order $Y$-shaped breather solution as shown in Fig.\ref{fig4} by choosing the appropriate parameters. Both the first and second components contain two eye-shaped wings and one anti-eye-shaped wing. The third component have three eye-shaped wings.

\section{Conclusions and discussions}\label{5s}
In conclusion, we have systematically investigated the defocusing TCKE system. The following key conclusions are drawn:

We construct binary DT for the defocusing three-coupled system based on the $4\times4$ Lax pair. The vector dark soliton is obtained through a limit technique. We perform an asymptotic analysis for the dark soliton component and validate the reliability of the asymptotic analysis method by comparing the asymptotic solution and the exact solution. Then we conduct a matrix analysis to obtain the breather solution and Y-shaped breather solution, which do not exist in the single-component defocusing KE system.  

This work highlights the effectiveness of binary DT in solving dark soliton solutions and deepen our understanding of mutual coupling effects in multi-component system. It will offer an insight into subsequent studies on defocusing multi-component systems.


\section*{Data availability}
Data sharing is not applicable to this article as no new data were
created or analyzed in this study.

\section*{Funding}
The authors have not disclosed any funding.
\section*{Author Declarations}
The author declare that there is no conflict of interest regarding the publication of this paper.

\section*{Appendix}\label{app:A}
\renewcommand{\theequation}{A.\arabic{equation}}
\setcounter{equation}{0} 
The undetermined parameters in the section \ref{2s} are exhibited as follows.
\begin{align*}
    &A_{11}=\frac{2}{2+\sqrt{2}\mathrm{i}}, A_{12}=\frac{\sqrt{2}}{2\mathrm{i}-\sqrt{2}},
    C_{11}=\frac{3}{3+\sqrt{7}\mathrm{i}},C_{12}=\frac{\sqrt{7}}{3\mathrm{i}-\sqrt{7}},\\
    &B_{11}=
\frac{4 \left( 3\mathrm{i}\sqrt{14} + 7\sqrt{2} + 12\mathrm{i} + 4\sqrt{7} \right)}{\left( 2\sqrt{2} + \sqrt{7} + \mathrm{i} \right) \left( -\frac{\sqrt{2}}{2} + \mathrm{i} \right) \left( \sqrt{7}\mathrm{i} + 3 \right) \left( 2\sqrt{2} + \sqrt{7} - \mathrm{i} \right)}, \\
&B_{12}=
\frac{\left( (16\sqrt{7} + 48\mathrm{i})\sqrt{2} + 24\mathrm{i}\sqrt{7} + 56 \right)}{\left( -2\sqrt{2} - \sqrt{7} + \mathrm{i} \right) \left( \sqrt{7}\mathrm{i} + 3 \right) \left( 2\sqrt{2} + \sqrt{7} + \mathrm{i} \right) \left( \sqrt{2}\mathrm{i} + 2 \right)},\\
&D_{11}=
\frac{72\sqrt{2}\sqrt{7} + 288}{\left( \sqrt{7}\mathrm{i} + 3 \right) \left( 2\sqrt{2} + \sqrt{7} + \mathrm{i} \right) \left( \sqrt{2}\mathrm{i} + 2 \right)^2 \left( 2\sqrt{2} + \sqrt{7} - \mathrm{i} \right)},\\
&D_{12}=
\frac{8 \left( 4\sqrt{7} + 7\sqrt{2} \right)}{\left( 2\sqrt{2} + \sqrt{7} + \mathrm{i}\right) \left( \sqrt{2}\mathrm{i} + 2 \right)^2 \left( -\frac{\sqrt{7}}{3} + \mathrm{i} \right) \left( 2\sqrt{2} + \sqrt{7} - \mathrm{i}\right)},\\
&A_{21}=0,A_{22}=-1,C_{21}=\frac{1}{1-\sqrt{7}\mathrm{i}},C_{22}=-\frac{\sqrt{7}}{\sqrt{7}+\mathrm{i}},D_{22}=-1,\\
&B_{21}=0,B_{22}=
\frac{4\mathrm{i}\sqrt{14} + 16\mathrm{i} - 28\sqrt{2} - 16\sqrt{7}}{\left( \sqrt{7} + \mathrm{i} \right) \left( 2\sqrt{2} + \sqrt{7} + \mathrm{i} \right) \left( -2\sqrt{2} - \sqrt{7} + \mathrm{i} \right)},\\
&D_{21}=
-\frac{4 \left( \sqrt{7}\mathrm{i} - 7 \right)\sqrt{2}}{\left( 2\sqrt{14} - 7 \right) \left( 2\sqrt{2} + \sqrt{7} + \mathrm{i} \right) \left( 2\sqrt{2} + \sqrt{7} - \mathrm{i}\right) \left( \sqrt{7} + \mathrm{i} \right)},\\
&A_{31}=\frac{1}{1-\sqrt{2}\mathrm{i}},A_{32}=-\frac{\sqrt{2}}{\sqrt{2}+\mathrm{i}},C_{31}=\frac{3}{3-\mathrm{i}\sqrt{7}},C_{32}=-\frac{\sqrt{7}}{3\mathrm{i}+\sqrt{7}},\\
&B_{31}=
\frac{-12\sqrt{14} - 16\mathrm{i}\sqrt{7} - 28\mathrm{i}\sqrt{2} - 48}{\left( 2\sqrt{2} + \sqrt{7} - \mathrm{i} \right) \left( 2\sqrt{2} + \sqrt{7} + \mathrm{i} \right) \left( \sqrt{7} + 3\mathrm{i} \right) \left( \sqrt{2} + \mathrm{i} \right)},\\
&B_{32}=
\frac{-16\mathrm{i}\sqrt{2}\sqrt{7} - 56\mathrm{i} - 48\sqrt{2} - 24\sqrt{7}}{3 \left( \frac{\sqrt{7}}{3} + \mathrm{i} \right) \left( -2\sqrt{2} - \sqrt{7} + \mathrm{i} \right) \left( 2\sqrt{2} + \sqrt{7} + \mathrm{i} \right) \left( \sqrt{2}\mathrm{i} - 1 \right)},\\
&D_{31}=
\frac{-36\sqrt{14} - 144}{\left( \sqrt{7}\mathrm{i} - 3 \right) \left( 2\sqrt{2} + \sqrt{7} + \mathrm{i} \right) \left( \sqrt{2} + \mathrm{i} \right)^2 \left( -2\sqrt{2} - \sqrt{7} + \mathrm{i} \right)},\\
&D_{32}=
\frac{-48\sqrt{7} - 84\sqrt{2}}{\left( \sqrt{7} + 3\mathrm{i} \right) \left( 2\sqrt{2} + \sqrt{7} + \mathrm{i} \right) \left( \sqrt{2} + \mathrm{i} \right)^2 \left( -2\sqrt{2} - \sqrt{7} + \mathrm{i} \right)}.
\end{align*}

\bibliographystyle{elsarticle-num} 
\bibliography{references} %

\end{document}